\def\ergps{{\rm\,erg\,s^{-1}}}

\def\kms{\ifmmode{\,{\rm km}\,{\rm s}^{-1}}\else {\rm\,km\,s$^{-1}$}\fi}

\def\msun{{\rm\,M_\odot}}

\def\kmps{{\rm\,km\,s^{-1}}}
\def\hmpc{\ifmmode{h^{-1}\,\hbox{Mpc}}\else{$h^{-1}$\thinspace Mpc}\fi}

\def\et{{\it et~al.}~}

\def\r200{\ifmmode{r_{200}}\else {$r_{200}$}\fi}

\def\omzeroc{0.19\pm0.06}

\documentstyle[11pt,aaspp4]{article}
\tighten

\begin{document}

\title
{Redshift Evolution of Galaxy Cluster Densities}

\author{
R.~G.~Carlberg\altaffilmark{1,2},
S.~L.~Morris\altaffilmark{1,3},
H.~K.~C.~Yee\altaffilmark{1,2},
\&
E.~Ellingson\altaffilmark{1,4},
}

\altaffiltext{1}{Visiting Astronomer, Canada--France--Hawaii Telescope, 
	which is operated by the National Research Council of Canada,
	le Centre National de Recherche Scientifique, and the University of
	Hawaii.}
\altaffiltext{2}{Department of Astronomy, University of Toronto, 
	Toronto ON, M5S~3H8 Canada}
\altaffiltext{3}{Dominion Astrophysical Observatory, 
	Herzberg Institute of Astrophysics,	
	National Research Council of Canada,
	5071 West Saanich Road,
	Victoria, BC, V8X~4M6, Canada}
\altaffiltext{4}{Center for Astrophysics \& Space Astronomy,
	University of Colorado, CO 80309, USA}

\begin{abstract}
The number of rich galaxy clusters per unit volume is a strong
function of $\Omega$, the cosmological density parameter, and
$\sigma_8$, the linear extrapolation to $z=0$ of the density contrast
in 8\hmpc\ spheres. The CNOC cluster redshift survey provides a sample
of clusters whose average mass profiles are accurately known, which
enables a secure association between cluster numbers and the filtered
density perturbation spectrum.  We select from the CNOC cluster survey
those EMSS clusters with bolometric $L_x\ge 10^{45}\ergps$ and a
velocity dispersion exceeding 800\kms\ in the redshift ranges
0.18-0.35 and 0.35-0.55.  We compare the number density of these
subsamples with similar samples at both high and low redshift.  Using
the Press-Schechter formalism and CDM style structure models, the
density data are best described with $\sigma_8\simeq0.75\pm0.1$ and
$\Omega\simeq0.4\pm0.2$ (90\% confidence). The cluster dynamical
analysis gives $\Omega=0.2\pm0.1$ for which $\sigma_8=0.95\pm0.1$
(90\% confidence).  The predicted cluster density evolution in an
$\Omega=1$ CDM model exceeds that observed by more than an order of
magnitude.
\end{abstract}

\keywords{galaxies: clusters, cosmology: large-scale structure of universe}

\section{Introduction}

The clustering of galaxies grows via gravity from density
perturbations which are characterized by their power spectrum, $P(k)$.
Various theories predict the shape of $P(k)$, but they do not
accurately predict its amplitude.  An integral constraint on the
normalization is conventionally parameterized as $\sigma_8$, the
fractional mass variance in 8 \hmpc\ spheres calculated using the
linear extrapolation of $P(k)$.  Rich galaxy clusters are particularly
sensitive probes of $\sigma_8$. N-body simulations have established
that the Press-Schechter formalism (\cite{ps}, hereafter PS) gives a
remarkably accurate prediction of the number of clusters per unit
cosmological volume as a function of mass.  Modeling the low redshift
data with the Press-Schechter formula, and using CDM style $P(k)$,
leads to a range of possibilities, from $\Omega=1$ and
$\sigma_8\simeq0.5$ to $\Omega\simeq0.2$ and $\sigma_8\simeq1$ and
values that interpolate between the two (\cite{ha,wef,ecf,vl,bm}).
The $\Omega$ dependence can be disentangled from $\sigma_8$ if data
giving $n(M)\,dM$ are available as a function of redshift (\cite{ob}).

The Canadian Network for Observational Cosmology (CNOC) cluster sample
and observational strategy (\cite{yec}) was specifically designed to
produce data useful for a $\sigma_8$ measurement.  The sample's
primary advantage is that the cluster masses are accurately known near
the virial radius which is essential for a reliable estimate of the
linear mass scale from which the cluster collapsed.  Here we combine
our results with similarly selected clusters at higher and lower
redshifts in Section3. The data are modeled in Section 4 to draw
conclusions about the values of $\sigma_8$ and $\Omega$.

\section{Press-Schechter Predictions}

The number density of clusters in the mass range $M$ to $M+dM$ is
predicted to be
\begin{equation}
n(M)dM = {{-3\delta_c(z)}\over{(2\pi r_L^2)^{3/2}\Delta}}
	 {{d\ln{\Delta}}\over{d\,M}} \exp{[-\delta_c^2(z)/2\Delta^2]}
	\,dM,
\label{eq:ps}
\end{equation}
where $\delta_c(z)=\delta_0(\Omega)/D(z,\Omega)$ gives the linear
overdensity at which a collapsed structure is approximately
virialized.  The function $\delta_0=0.15(12\pi)^{2/3}\Omega^{0.0185}$
is nearly constant at $\delta_0\simeq 1.68$ (\cite{nfw}).  The growth
factor, $D(z,\Omega)$, gives the redshift dependence of the linear
amplitude of the density perturbations (\cite{ppc}).  The quantity
$\Delta(r_L)$ measures the fractional linear mass variance in spheres
of radius $r_L$.  It is calculated using a tophat filter from a
parameterized version of the CDM spectrum ($\Gamma$ fixed at 0.2
\cite{ebw}) whose normalization is adjusted such that
$\sigma_8=\Delta(8\hmpc)$.  The ``just virialized'' radius is near
1.5\hmpc\ and has an overdensity of $\simeq178\Omega^{-0.6}$
(\cite{wef}).  Near 1.5\hmpc\ $M(r)\propto r^p$, where $p\simeq0.64$
for the CNOC data (\cite{profile}).  The tophat filtering scale,
$r_L$, is related to $M_{1.5}$, as $r_L \simeq 8.43
\Omega(z)^{0.2p/(3-p)} [M_{1.5}/6.97\times10^{14} h^{-1} \msun/
\Omega]^{1/(3-p)} (1+z)^{-p/(3-p)}\hmpc$ (a generalization of
\cite{wef}) where $\Omega(z)$ is the value of $\Omega$ at redshift $z$.

We integrate Eq.~\ref{eq:ps} from the minimum mass in the sample to
infinity to derive $n(>M_{1.5})$.  For comparison with measurements,
we average the density predictions over the redshift ranges of
interest.  These predictions are plotted in Figures~\ref{fig:no} and
\ref{fig:nc} for $\Omega=0.2$ and $\Omega=1$, respectively.

\section{Cluster Density Estimates}

The CNOC sample was drawn from the EMSS cluster survey (\cite{emss}),
supplemented with a few high redshift clusters identified later
(\cite{lg}). We impose the constraints $f_x \ge
4\times10^{-13}\,$erg~cm$^{-2}$~s$^{-1}$, $L_x({\rm 0.3-3.5\,Kev}) \ge
4\times 10^{44} \ergps$, redshifts between 0.18 and 0.55, and in the
declination range $-15$ to $+65$ degrees. We have recalculated the
$f_x$ and $L_x$ to better allow for the redshift dependence of the
flux in the ``detect cell''. The main difference is that the cluster
MS0302+16 now falls below our luminosity cut, although this has a
negligible affect on the results (see also \cite{nichol}).

The $M_{1.5}$ masses for each cluster in the CNOC sample are estimated
to be $ M_v b_{Mv}\, (1.5\hmpc/r_v)^p $ where $b_{Mv}$ is the
virial mass bias, $r_v$ is the virial radius, calculated as a ringwise
potential (\cite{global}), and $M_v$ is the resulting virial mass.
For the CNOC clusters $b_{Mv}\simeq 0.85$ on the average
(\cite{profile}), that is, the virial mass is always an
overestimate. The masses $M_{1.5}$ are calculated from the $M_v$ and
$r_v$ of the CNOC clusters (\cite{global} as updated slightly by the
use of the finalized catalogues).

\subsection{Correcting for X-Ray Selection}

We need to count the number of clusters having $M_{1.5}$ larger than a
specified mass within some redshift range.  A significant complication
is that we have a sample defined by its X-ray properties, but we want
to make a measurement based on its distribution of characteristic
masses. It is well known that there is a strong $L_x-M_{1.5}$
correlation, but with a substantial scatter (for instance, \cite{es},
hereafter ES).  That is, some clusters that are in the $L_x$ limited
sample will be below the specified mass limit and vice versa.  To
correct from an X-ray selected sample to a mass selected sample we
proceed as follows.  We use the Edge and Stewart study of the X-ray
and optical properties of nearby clusters, which appear to be
essentially identical to the CNOC sample in the relevant parameters,
see Figure~\ref{fig:lxsig}.  The ES sample is effectively a mass
selected sample, at least for the rich clusters which are our concern
here. All the selected clusters have a bolometric $L_x\ge
10^{45}\ergps$.  Examining Figure~\ref{fig:lxsig}, we see that setting
the minimum mass equal to that for a cluster with $\sigma_v
\simeq 800
\kms$ will help maximize the number of clusters in the sample while keeping
the corrections from the $L_x$ selected sample to the $\sigma_v$
selected sample relatively small.  

There are two different correction factors for $L_x$ defined samples,
depending on whether or not the parent sample has known velocity
dispersions.  In the ES sample there are 12 clusters with $\sigma_v\ge
800\kms$, of which 6 are also above $L_x\ge 10^{45}\ergps$. Therefore
we estimate that for a sample selected with both the $L_x$ and
$\sigma_v$ limits, the true density of clusters with $\sigma_v\ge
800\kms$ is a factor of $f_{x\sigma}=12/6=2\pm1$ times the density
measured.  There are 8 ES clusters with $L_x\ge 10^{45}\ergps$ for all
velocity dispersions and 12 with $\sigma_v\ge 800\kms$.  Therefore for
a sample selected with only the $L_x$ limit, we estimate that the
number of clusters above $\sigma_v=800\kms$ is $f_x=12/8=1.5\pm0.7$
times the measured density.  Although these corrections are not very
accurate, they illustrate that the corrections are not large, and
their errors are comparable in size to those from the subsamples
themselves.

\subsection{The CNOC Sample}

To convert from EMSS 0.3-3.5Kev luminosities to the bolometric
luminosity we derive a mean bolometric correction of a factor of 2.20
at 6.8~Kev and 3.12 at 13.6~Kev, with only a small dependence on the
HI column. We adopt a uniform bolometric correction of a factor of 2.5
for the CNOC clusters (see the high $L_x$ sample of
\cite{hjg} for representative EMSS cluster temperatures).  
Imposing the limit $\sigma_v\ge 800 \kms$ reduces the CNOC sample size
from 12 clusters to 8 (2 of the 16 clusters observed being dropped as
being below the X-ray limits, 1 cluster not from EMSS, and 1 has a
poorly determined $\sigma_v$).  Using the $V_e/V_{max}$ method
(\cite{ab}) we then measure the volume density of these clusters,
allowing for the EMSS sky area in the CNOC region as a function of
flux (see \cite{emss} for more details on this procedure and also an
example table of EMSS all-sky coverage).  The mean densities of these
clusters for $\Omega=0.2$ are reported in Table~\ref{tab:deno}.  The
CNOC sample is split into a low redshift subsample, $0.18\le z \le
0.35$, which has a smallest $M_{1.5}$ of $4.8\times10^{1}h^{-1}\msun$
and a moderate redshift subsample, $0.35\le z \le 0.55$, which is
found to have a smallest $M_{1.5}$ of
$6.7\times10^{14}h^{-1}\msun$. The $f_{x\sigma}$ corrected densities
are given in Table~\ref{tab:deno}.

\subsection{Low-Redshift Samples}

At low redshift there are several samples to consider as sources of
density estimates.  The most straightforward dataset is that of Henry
\& Arnaud (1991, hereafter HA) who provide an X-ray luminosity function at 
low redshift, which when integrated from $L_x=10^{45} \ergps$ to
infinity yields an ($f_x$ corrected) volume density of
$\simeq7.5\times 10^{-7} h^3$\,Mpc$^{-3}$.  The velocity dispersions
of these clusters extend below 800 \kms, but none below 750 \kms
(\cite{zhg}), which we adopt as the minimum velocity dispersion. We
scale the cluster masses with velocity dispersion as $M\propto
\sigma_v^3$ with a reference value of $5.7\times10^{14}h^{-1}\msun$.  
We estimate the $M_{1.5}$ of the HA
sample as $4.7\times 10^{14}h^{-1}\msun$.

The ESO Cluster Survey (\cite{eso}) finds a cluster density of
$2.5\times 10^{-6} h^3$~Mpc$^{-3}$ for $\sigma_v\ge 800\kms$ at
$z\le0.1$. However, the ESO and CNOC velocity dispersions are not
calculated in the same manner.  The CNOC velocity dispersions are
estimated using an explicit background subtraction.  On the average we
find that our velocity dispersions are about 7\% lower than those
calculated from precisely the same velocities using the iterated
bi-weight estimator (\cite{bfg,profile}). As we increase the redshift
range of the data given to the bi-weight estimator up to 25\%, the
velocity dispersion rises an average of 13\% and then remains
reasonably stable. We adjust the ESO velocity dispersions downwards by
13\%, to arrive a velocity dispersion of 708 \kms.  We derive a mass
limit for the ESO $\ge 800 \kms$ sample of
$M_{1.5}=4.0\times10^{14}h^{-1}\msun$.  The sample is not X-ray selected so
needs no density correction.

An upper limit to the low redshift density for the Northern Abell
sample with velocity dispersions has been derived previously
(\cite{wef}). The same sample, when compared with X-ray results
confirms that the median velocity dispersion is somewhat overestimated
(\cite{ecf} referred to as ECF), and argues that the velocity
dispersion is about 650
\kms.  Because this sample is very similar to the ESO sample, we adopt
the same minimum velocity dispersion as we derived above, 708 \kms,
and hence the same minimum mass. This sample requires no density
correction.

\subsection{A High-Redshift Sample}

The EMSS sample contains a fair sample of clusters ranging from
redshifts of about 0.14 to redshift 0.83 (\cite{emss,lg}).  The high
redshift sample (\cite{lg}) is cautiously assigned the same minimum
mass, $M_{1.5}=4.7\times 10^{14} h^{-1}\msun$ as we used for the HA
subsample. Given the richness of these clusters we expect that the
clusters actually have higher masses, consequently this is a
conservative assumption. The mean density derived from the 4 clusters
in the redshift range $0.55 \le z \le 0.85$ for $\Omega=0.2$ and 1 is
given in the Table and includes the $f_x$ correction. Reassuringly,
they are similar to the densities derived elsewhere (\cite{lg}) using
the same clusters, but a different analysis.  It is known that at
least one of these clusters, MS1054$-$03, likely has a mass well over
the limit to be included in the sample (\cite{lk}). The minimum mass
is uncertain but likely at least that of the moderate $z$ CNOC sample.
In particular, these are very rich clusters which the very strong
richness-$\sigma_v$ relation (\cite{global}) indicates to be high
$\sigma_v$ clusters. We expect $\sigma_v\simeq 900 \kms$.

\section{Parameter Probabilities and Conclusions}

The $\chi^2$ probability contours of the PS predictions and the
observed densities are plotted in Figure~\ref{fig:chi2}. The sample
variances are calculated as the quadrature sum of the $1/\sqrt{N}$ of
the sample and the errors in the sample correction factors.  The
observed densities as a function of $\Omega$ are estimated as a linear
interpolation between the values at $\Omega=0.2$ and $\Omega=1$
measurements.  The mass errors are known to be about 25\% for a single
cluster (\cite{global}), which in the mean for the various samples
will be reduced to about 8-15\%, depending on sample size. This
relatively small error is neglected in calculating $\chi^2$.

The minimum $\chi^2$ is near $\Omega\simeq0.4$ and
$\sigma_8\simeq0.75$, although $\Omega$ between 0.2 and 0.6 are
statistically acceptable within the 90\% confidence interval.  For our
preferred $\Omega=0.2$ we find that $\sigma_8=0.95\pm0.05$. A model
with $\Omega=1$ is excluded at more than 99\% confidence, although
this is dependent on the the estimated masses of the high redshift
EMSS sample.

\acknowledgments
We thank Mike Hudson, Jim Bartlett, Alain Blanchard and Jamila Oukbir
for comments.  The Canadian Time Assignment Committee allocated CFHT
observing time. The CFHT organization provided the technical support
which made these observations feasible.  Funding was provided by NSERC
and NRC of Canada.

\clearpage
\begin{table}[h]
\caption{Cluster Densities for $\Omega=0.2$\label{tab:deno}}
\smallskip
\begin{tabular}{rrrrr}
\tableline
\smallskip
redshift & Sample & $\sigma_v$(min) & $\log{M}$ & $\log{n(>M)}$ \\
 & & $\kmps$ & $h^{-1}\msun$  & h$^{-3}$\,Mpc$^3$ \\
\tableline
0.0-0.10 & WEF & 850 & 14.66 & -5.40 \\
0.0-0.10 & ECF & 708 & 14.60 & -5.40 \\
0.0-0.10 & ESO & 708 & 14.60 & -5.60 \\
0.0-0.10 & HA  & 750 & 14.67 & -6.12 \\
0.18-0.35 &CNOC& 800 & 14.68 & -6.53 \\
0.35-0.55 &CNOC& 800 & 14.80 & -7.15 \\
0.55-0.85 &EMSS& 800 & 14.81 & -7.30 \\
\end{tabular}
\end{table}

\newpage

\figcaption{
The $n(>M)$ relation for the different redshift ranges as calculated
from the PS formula for $\Omega=0.2$ and $\sigma_8=0.95$. The errors
are $N^{-1/2}$ estimates with the selection bias errors added in
quadrature.  Optically selected clusters have open symbols, X-ray
selected clusters have closed symbols. The CNOC and EMSS samples have
densities derived in this paper. The low redshift samples are
discussed in the text.
\label{fig:no}}
\begin{figure}[h]\figurenum{1}\plotone{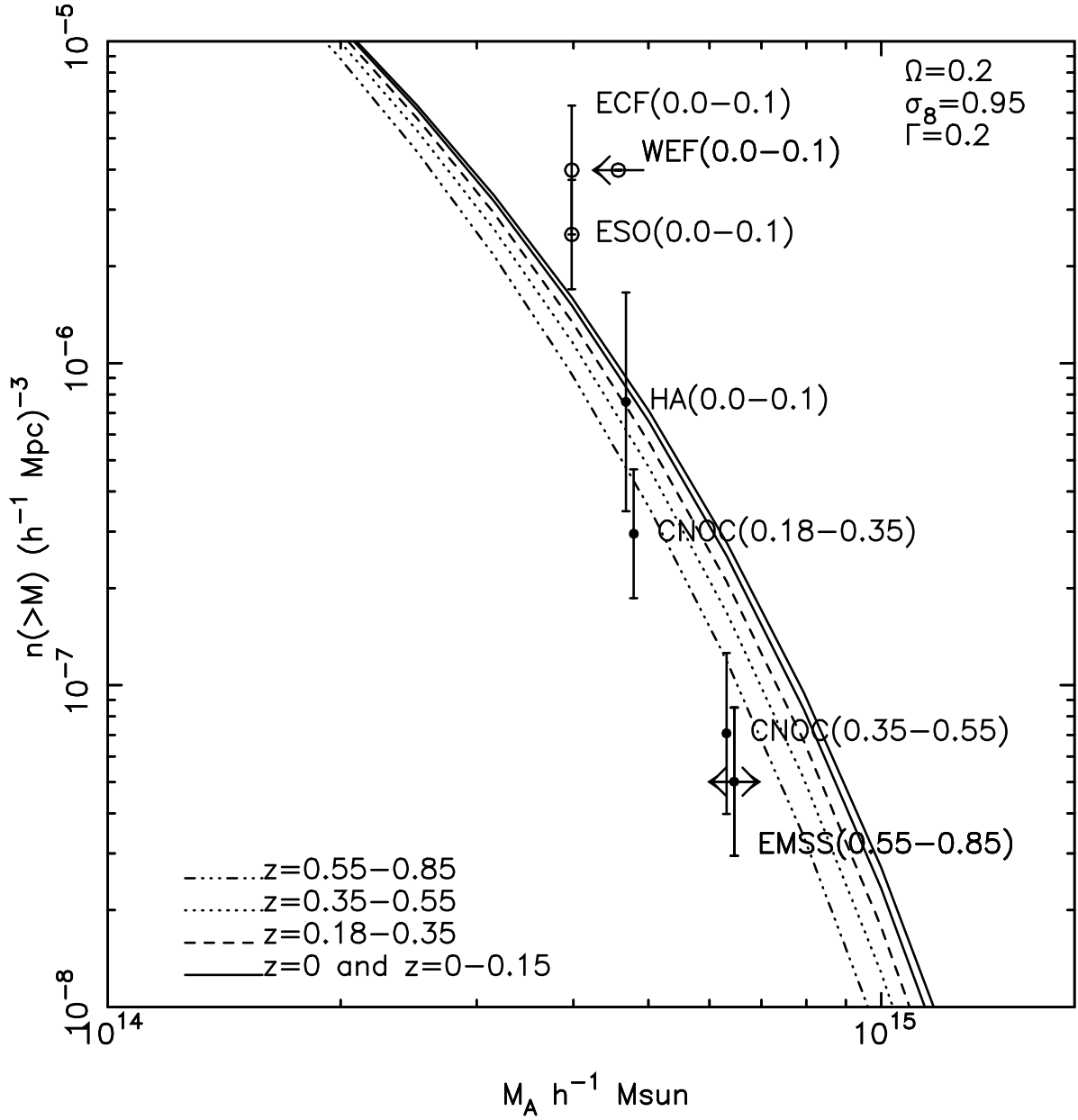}
\caption{
The $n(>M)$ relation for the different redshift ranges as calculated
from the PS formula for $\Omega=0.2$ and $\sigma_8=0.95$. The errors
are $N^{-1/2}$ estimates with the selection bias errors added in
quadrature.  Optically selected clusters have open symbols, X-ray
selected clusters have closed symbols. The CNOC and EMSS samples have densities
derived in this paper. The low redshift samples are discussed in the
text.
}\end{figure}

\figcaption{
The $n(>M)$ relation from the PS formula calculated for $\Omega=1$ and
$\sigma_8=0.55$. These model parameters do not adequately describe the data.
\label{fig:nc}}
\begin{figure}[h]\figurenum{2}\plotone{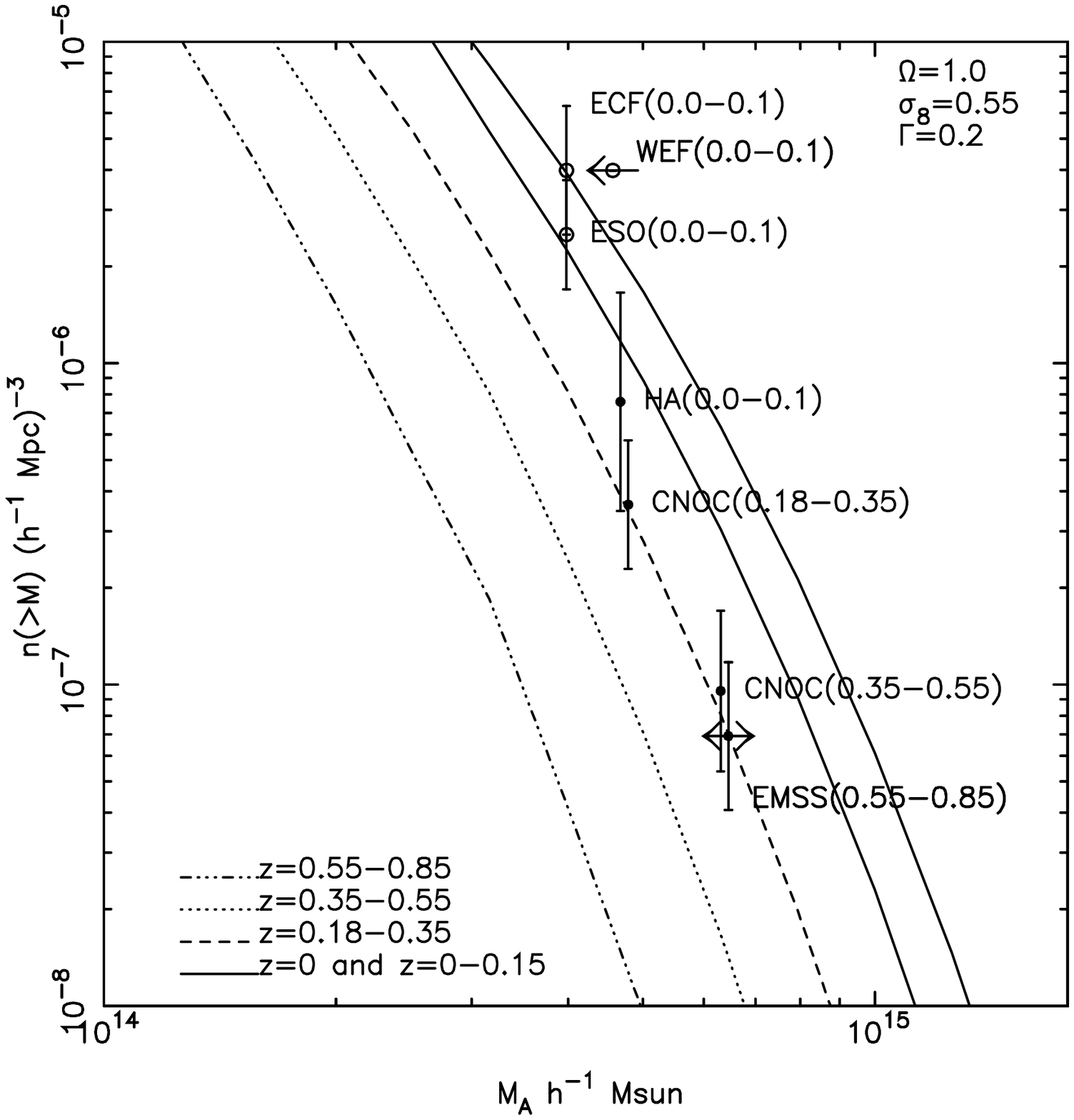}
\caption{
The $n(>M)$ relation from the PS formula calculated for $\Omega=1$ and
$\sigma_8=0.55$. These model parameters do not adequately describe the data.
}\end{figure}

\figcaption{
The bolometric X-ray luminosity versus line-of-sight
velocity dispersion for the Edge \& Stewart and CNOC samples.
The sample limits at $L_x\ge 10^{45}\ergps$ and $\sigma_v\ge 800 \kms$
are shown. The ES velocity dispersions are divided by 1.13.
\label{fig:lxsig}}
\begin{figure}[h]\figurenum{3}\plotone{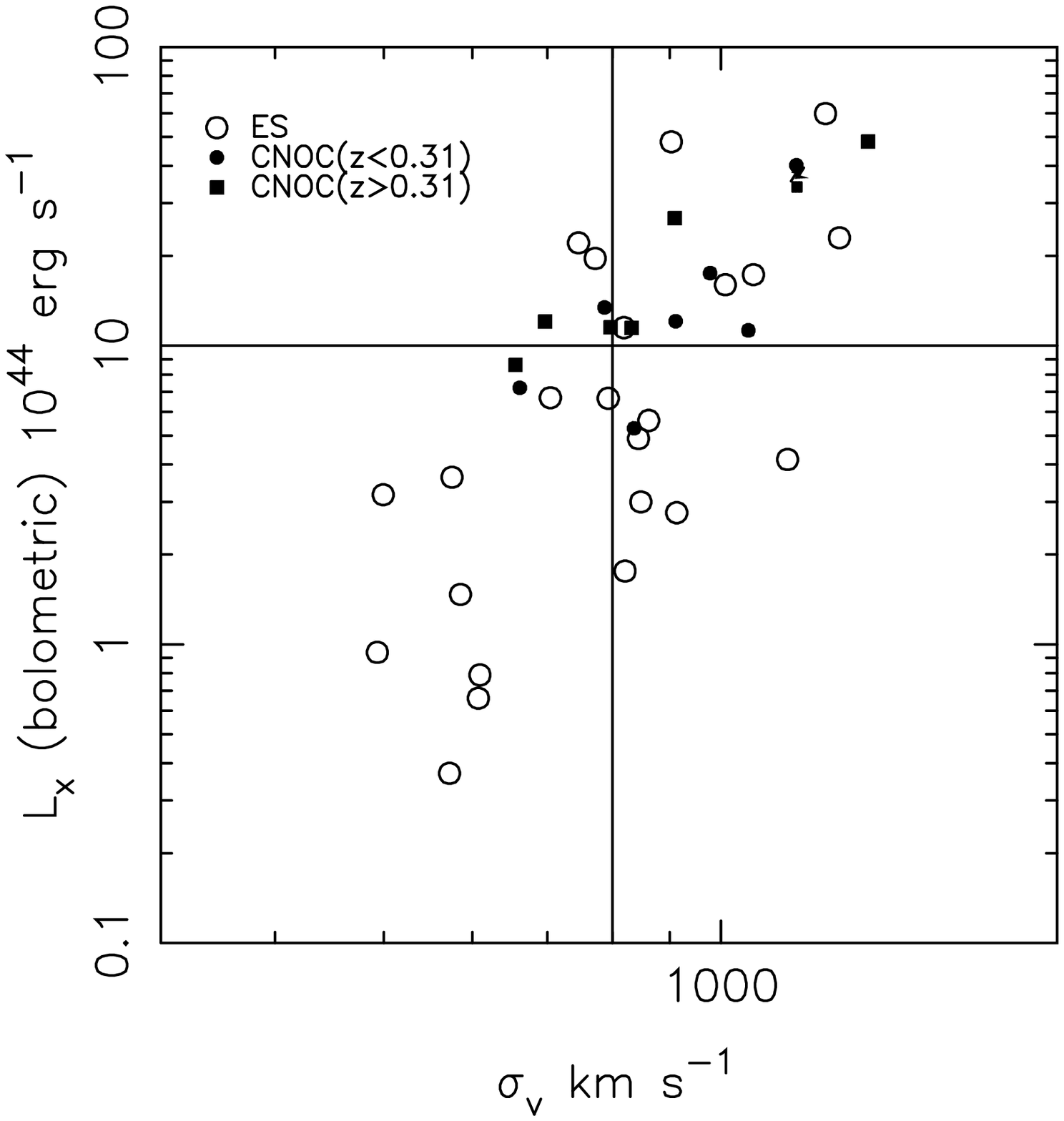}
\caption{
The bolometric X-ray luminosity versus line-of-sight
velocity dispersion for the Edge \& Stewart and CNOC samples.
The sample limits at $L_x\ge 10^{45}\ergps$ and $\sigma_v\ge 800 \kms$
are shown. The ES velocity dispersions are divided by 1.13.
}\end{figure}

\figcaption{
A plot of $\chi^2$ for the all {\em independent} samples (solid lines)
and excluding the high redshift EMSS sample (dotted lines). The
contours are the 90\% and 99\% confidence levels.  The results of the
CNOC analysis, $\Omega=\omzeroc$, with its $1\sigma$ range are
indicated.
\label{fig:chi2}}
\begin{figure}[h]\figurenum{4}\plotone{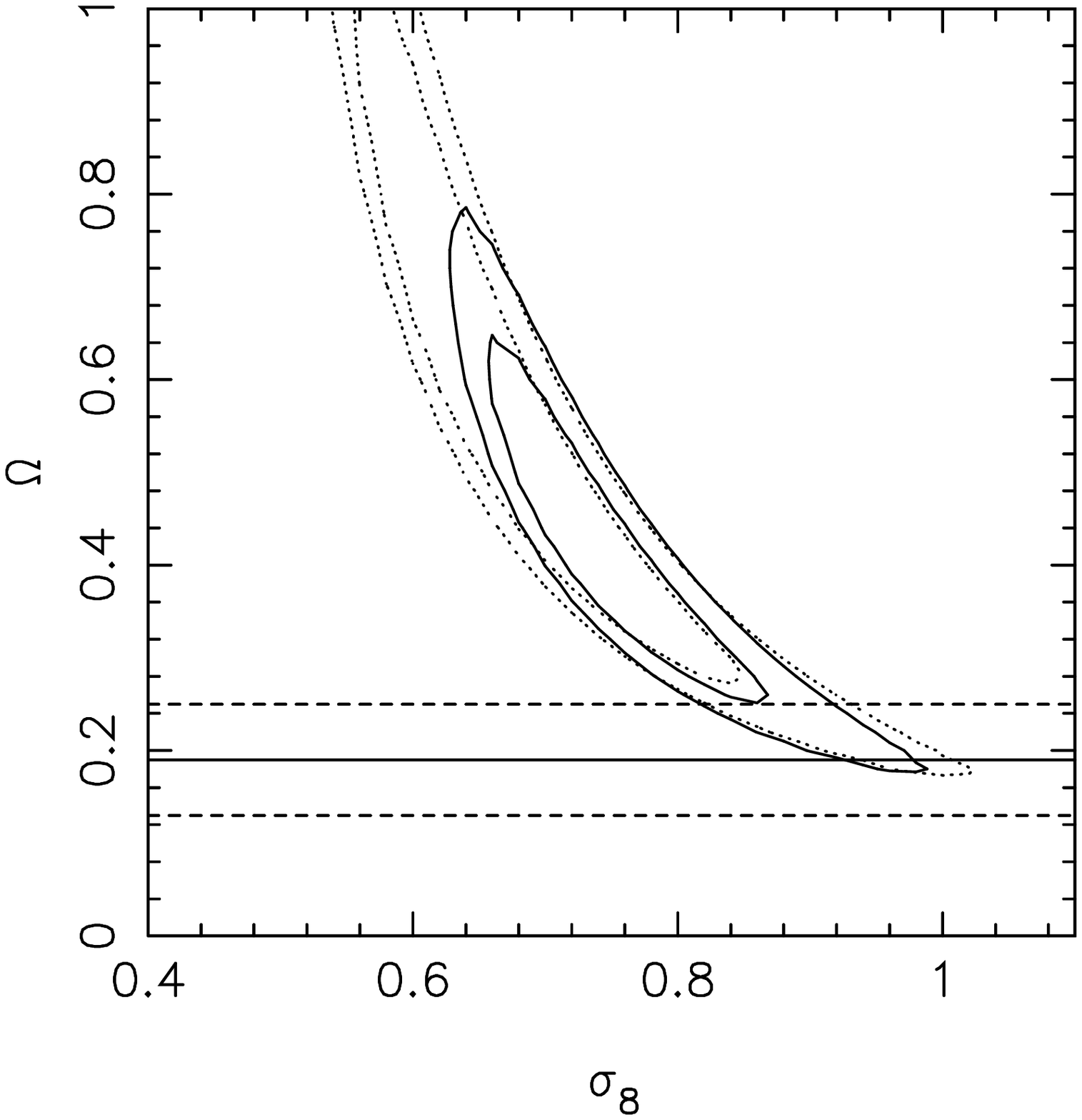}
\caption{
A plot of $\chi^2$ for the all {\em independent} samples (solid lines)
and excluding the high redshift EMSS sample (dotted lines). The
contours are the 90\% and 99\% confidence levels.  The results of the
CNOC analysis, $\Omega=\omzeroc$, with its $1\sigma$ range are
indicated.
}\end{figure}

\end{document}